\documentclass[twocolumn]{svjour3}         
\smartqed  
\usepackage{graphicx}
\usepackage{textcomp}
\usepackage{amssymb}
\usepackage{wasysym}

 \journalname{Nanoscale Research Letters}
\begin{document}

\title{CMOS compatible dense arrays of Ge quantum dots on the Si(001) surface
}
\subtitle{Hut cluster nucleation, atomic structure, and array life cycle during UHV MBE growth}

\titlerunning{CMOS compatible dense arrays of Ge quantum dots on the Si(001) surface}        

\author{L.~V.~Arapkina   \and
        V.~A.~Yuryev       
}


\institute{Larisa~V.~Arapkina \at
              A. M. Prokhorov General Physics Institute of RAS, 38 Vavilov Street, Moscow, 119991, Russia \\
              Tel.: +7-499-5038318\\
              \email{arapkina@kapella.gpi.ru}           
           \and
           Vladimir~A.~Yuryev \at
              Tel.: +7-499-5038144\\
              Fax: +7-499-1350356\\
              \email{vyuryev@kapella.gpi.ru} 
}

\date{Received: date / Accepted: date}

\maketitle

\begin{abstract}
We report a direct observation of Ge hut nucleation on Si(001) during UHV MBE at $360\,^\circ$C. Nuclei of pyramids and wedges were observed on the  wetting layer 
$(M\times N)$ patches starting from the coverage of $>5.1$ \r{A} and found to have different structures. Atomic models of nuclei of both hut species have been built as well as models of the growing clusters. The growth of huts of each species has been demonstrated to follow generic scenarios. The formation of the second atomic layer of a wedge results in rearrangement of its first layer. Its ridge structure does not repeat the nucleus. A pyramid grows without phase transitions. A structure of its vertex copies the nucleus. Transitions between hut species turned out to be impossible. The wedges contain point defects in the upper corners of the triangular faces and have preferential growth directions along the ridges. The derived structure of the \{105\} facet follows the paired dimer (PD) model. Further growth of hut arrays results in domination of wedges, the density of pyramids exponentially drops. The second generation of huts arises at coverages $>10$ \r{A}; new huts occupy the whole wetting layer at coverages $\sim 14$ \r{A}.   Nanocrystalline Ge 2D layer begins forming at coverages $>14$\,\r{A}.
\end{abstract}

\section{\label{sec:intro}Introduction}

Development of CMOS compatible processes of formation of  germanium quantum dot (QD) dense arrays   on the (001) silicon surface as well as multilayer Ge/Si epitaxial heterostructures on their basis is a challenging task of great practical significance \cite{Report_01-303,Pchel_Review,Wang-properties,Smagina,Wang-Cha,WG-near_IR,Photonic_crystal,QDFET,QDIP-Wang,QDIP-Wang1,Sabelnik-2,Dvur-IR-20mcm,QDIP-MIS,QDIP-Wang2}.
An important direction of applied researches in this area is the development of highly efficient monolithic far and  mid infrared  detector arrays which could be produced by a standard CMOS technology \cite{QDIP-Wang,QDIP-Wang1,Sabelnik-2,Dvur-IR-20mcm,QDIP-MIS,QDIP-Wang2}. Such detectors have to combine high perfection (uniformity, sensitivity, operating life, etc.)  with high yield and low production price. A requirement of CMOS compatibility of technological processes imposes   a hard constraint on conditions of all phases of the QD array manufacturing starting from the stage of preparation of a clean  Si surface for Ge/Si heterostructure deposition: on the one hand, formation of a photosensitive layer must be one of the latest operations of the whole  device production cycle because otherwise the structure with QDs would be destroyed  by further high temperature annealings; from the other hand, high temperature processes during Ge/Si heterostructure formation on the late phase of the detector chip production would certainly wreck the readout circuit  formed on the crystal. Therefore,  lowering of the array formation temperature down to the values of $\lesssim 450^\circ$C\footnote{As well as decreasing of the wafer annealing temperatures and times during the clean Si(001) surface preparation.}   is strongly required \cite{Report_01-303,Sabelnik-2}, and the Ge QD arrays meeting this requirement are referred to as CMOS compatible ones.

In addition to the requirement of the low temperature of a Ge QD array formation,  
 both high density of the germanium nanoclusters ($> 10^{11}$~cm$^{-2}$) and high uniformity of the cluster shapes and sizes (dispersion $<$ 10\,\%)   in the arrays are necessary for  employment of such structures  in CMOS IR detectors \cite{Dvur-IR-20mcm}.
The molecular beam epitaxy (MBE) is known to be the main technique of formation of Ge/Si heterostructures with QDs  \cite{Pchel_Review,Brunner}. A high density of the self-assembled hut clusters can be obtained in the MBE process of the Ge/Si(001) structure formation when depositing germanium on the Si(001) substrate heated to a  temperature $T_{\rm gr}\lesssim 550^\circ$C. In this case the lower  is the temperature of the silicon substrate during the Ge deposition the higher is the density of the clusters at the permanent quantity of the deposited Ge \cite{Yakimov,Jin}. For example, the density of the Ge clusters in the array was  $6\times 10^{11}$~cm$^{-2}$ at   $T_{\rm gr} = 360^\circ$C and the effective thickness of the deposited germanium layer\footnote{I.e. the Ge coverage or, in other words, the thickness of the Ge film measured by the graduated in advance film thickness monitor with the quartz sensor installed in the MBE chamber.} $h_{\rm Ge} = 8~{\rm\AA}$; the cluster density of only $\sim 2\times 10^{11}$\,cm$^{-2}$ was obtained at  $T_{\rm gr} = 530^\circ$C and the same value of $h_{\rm Ge}$  \cite{classification}.

There is another approach to obtaining  dense cluster arrays. The authors of Refs.~\cite{Smagina,Ion_irradiation_1,Dvur_irrad,Ion_irradiation}   reached the cluster density of  $\sim 9\times 10^{11}$\,cm$^{-2}$  using the pulsed irradiation of the substrate by a low-energy Ge$^+$ ion beam during the MBE growth of the Ge/Si(001) heterostructures at $T_{\rm gr}$ as high as $570^\circ$C.

Obtaining of the arrays of the densely packed Ge QDs on the Si(001) surface is an important task but the problem of formation of  uniform arrays of the Ge clusters  is much more challenging one. The process of Ge/Si(001) heterostructure formation with the Ge QD dense arrays and predetermined electrophysical and photoelectric parameters cannot be developed until both of these tasks are  solved. The uniformity of the cluster sizes and shapes in the arrays determines not only the widths of the energy spectra of the charge carrier bound states in the QD arrays  \cite{Smagina} but in a number of cases the optical and electrical properties of both the arrays themselves and the  device structures produced on their basis \cite{Electrolumin}. To find an approach to the improvement of the Ge QD array uniformity on the Si(001) surface it is necessary to carry out a detailed morphological investigation of them.

This article presents the results of our recent investigations of several important issues of the Ge dense array formation and growth. We have studied the array nucleation phase (the transition from 2D growth of the wetting layer (WL) to 3D formation of the QD array when the nuclei of both species of huts---pyramids and wedges \cite{classification}---begin to arise on the $(M\times N)$ patches of WL). We have identified by STM the nuclei of both species, determined their atomic structure \cite{classification,Hut_nucleation} and observed the moment of appearance the first generation of the nuclei. We have investigated with high spatial resolution the peculiarities of each species of huts and their growth and derived their atomic structures \cite{Hut_nucleation,atomic_structure}. We have concluded that the wedge-like huts form due to a phase transition reconstructing the first atomic step of the growing cluster when dimer pairs of its second atomic layer stack up; the pyramids grow without such phase transitions. In addition, we have come to conclusion  that wedges contain  vacancy-type defects on the penultimate terraces of their triangular facets \cite{Hut_nucleation} which may decrease the energy of addition of new atoms  to these facets and stimulate the quicker growth on them than on the trapezoidal ones and rapid elongation of wedges. We have shown also comparing the structures and growth of pyramids and wedges that shape transitions between them are very unlikely \cite{Hut_nucleation,atomic_structure}. Finally, we have explored the array evolution during MBE right up to the end of its life  when most of clusters coalesce and start forming a  nanocrystalline 2D layer. 

Below, we present these results in detail.

\section{Methods, equipment and conditions of experiments}
\label{sec:setup}

The experiments were made using an integrated ultrahigh vacuum instrument \cite{classification} built on the basis of the Riber surface science center with the  EVA~32 molecular beam epitaxy chamber  connected to the STM GPI-300 ultrahigh vacuum scanning  tunnelling microscope  \cite{gpi300,STM_GPI-Proc,STM_calibration}.
This equipment allows us to carry out the STM study of samples at any phase of a substrate surface preparation and MBE growth. The samples can be  transferred into the STM chamber for the examination and moved back into the MBE vessel for further processing  as many times as required never leaving the UHV ambient and preserving the required cleanness for STM investigations with atomic resolution and MBE growth.

Initial substrates  were 8$\times$8 mm$^{2}$ squares cut from the specially treated commercial B-doped    CZ Si$(100)$ wafers ($p$-type,  $\rho\,= 12~\rm\Omega\,$cm). 
After washing and chemical treatment following the standard procedure described elsewhere \cite{phase_transition_ru,STM_RHEED} (which included washing in ethanol, etching in the mixture of HNO$_3$ and HF and rinsing in the deionized water), the silicon substrates were 
 mounted on the molybdenum STM holders and inflexibly clamped with the tantalum fasteners. The STM holders were placed in the  holders for MBE made of molybdenum with tantalum inserts.  Then the substrates were 
 loaded into the  airlock and transferred into the preliminary annealing chamber where outgassed at  the temperature of around $565^\circ$C and the pressure of about  $5\times 10^{-9}$ Torr for about 24 hours. After that the substrates were moved for final treatment into the MBE chamber evacuated down to about $10^{-11}$\,Torr. There were two stages of annealing in the process of substrate heating in the MBE chamber\,--- at $\sim 600^\circ$C for $\sim 5$ minutes and at $\sim 800^\circ$C for $\sim 3$ minutes \cite{classification}. The final annealing at the temperature greater than $900^\circ$C was carried out for nearly $ 2.5$ minutes with the maximum temperature of about $ 925^\circ$C ($\sim 1.5$~minutes). Then the temperature was rapidly lowered to about $ 750\,^\circ$C. The rate of the further cooling  was around $0.4^\circ$C/s that corresponded to the ``quenching'' mode applied in \cite{STM_RHEED}. The pressure in the MBE chamber grew to nearly $2\times 10^{-9}$ Torr during the deoxidization process.
The surfaces of the silicon substrates were completely purified of the oxide film as a result of this treatment; more data on the morphology of the prepared Si(001) clean surfaces can be found in Refs.~\cite{phase_transition_ru,STM_RHEED,our_Si(001)_en}. 

\begin{figure}
\includegraphics[width=0.4\textwidth]{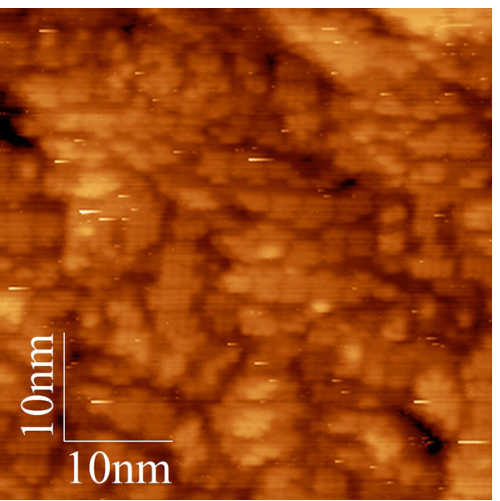}(a)
\includegraphics[width=0.4\textwidth]{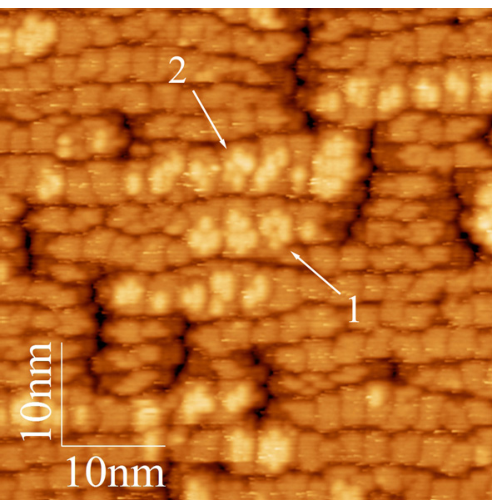}(b)
\caption{\label{fig:nucleation}
STM image of  Ge wetting layer on Si(001): (a) before cluster nucleation,  $h_{\rm Ge}=4.4$~\r{A}  ($U_{\rm s}=-1.86$~V, $I_{\rm t}=100$~pA); (b) arising nuclei of pyramidal (1) and wedgelike (2) huts,  $h_{\rm Ge}=5.1$~\r{A}  ($U_{\rm s}=+1.73$~V, $I_{\rm t}=150$~pA).}
\end{figure}

Ge was deposited directly on the deoxidized Si(001) surface from the source with the electron beam evaporation\footnote{The Si source was switched off during the experiments.}. The  Ge deposition rate was about $0.15$~\r{A}/s; the effective Ge film thickness $h_{\rm Ge}$ was varied from 4~\r{A} to 15~\r{A} for different samples. The deposition rate and  $h_{\rm Ge}$ were measured by the XTC film thickness monitor with the graduated in advance quartz sensor installed in the MBE chamber. The substrate temperature $T_{\rm gr}$ was $360^\circ$C during Ge deposition; the pressure in the MBE chamber did not exceed $10^{-9}$ Torr. The rate of the sample cooling down to the room temperature was approximately $0.4^\circ$C/s after the deposition. 

The samples were heated by Ta radiators from the rear side in both preliminary annealing and MBE chambers. The temperature was monitored with chromel-alumel and tungsten-rhenium thermocouples in the preliminary annealing and MBE chambers, respectively. The thermocouples were mounted in vacuum near the rear side of the samples and {\it in situ} graduated beforehand against the IMPAC~IS\,12-Si pyrometer which measured the sample temperature through chamber windows.
The atmosphere composition in the MBE camber was monitored using the SRS~RGA-200 residual gas analyzer before and during the process.

After Ge deposition and cooling, the prepared samples were moved for analysis into the STM chamber in which the pressure did not exceed $10^{-10}$ Torr. The STM tip was {\it ex situ} made of the tungsten wire and cleaned by ion bombardment \cite{W-tip} in a special UHV chamber connected to the STM one. The images were obtained in the constant tunneling current ($I_{\rm t}$) mode at the room temperature. The STM tip was zero-biased while the sample was positively or negatively biased ($U_{\rm s}$) when scanned in empty or filled states imaging mode.

Original firmware  \cite{gpi300,STM_GPI-Proc,STM_calibration} was used for data acquisition; the STM images were processed afterwords using the WSxM software \cite{WSxM}.

\section{\label{sec:results}Experimental data and structural models}
\subsection{\label{sec:formation}Array and hut cluster nucleation}

\begin{figure}
\includegraphics[width=0.4\textwidth]{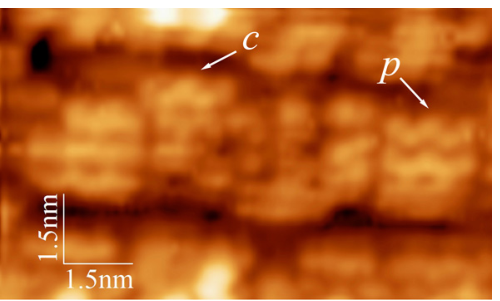}(a)
\includegraphics[width=0.4\textwidth]{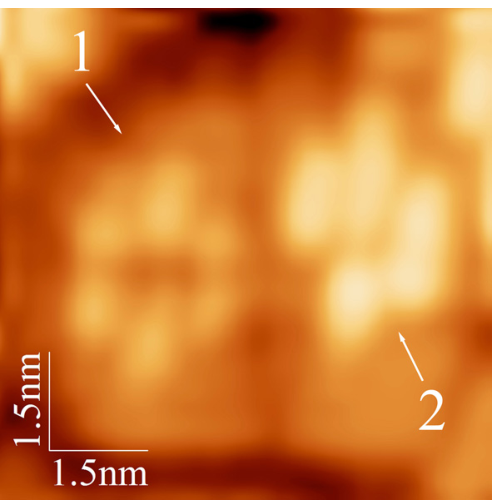}(b)
\caption{\label{fig:nuclei}
STM image of  Ge wetting layer on Si(001): (a) $c(4\times 2)$~$(c)$ and $p(2\times 2)$ $(p)$ reconstructions within the $(M\times N)$ patches,  $h_{\rm Ge}=6,0$~\r{A}, $U_{\rm s}= +1.80$~V, $I_{\rm t}=80$~pA; (b) new formations arise on the $(M\times N)$ patches   due to nucleation  of Ge pyramid (1) and wedge (2),  $h_{\rm Ge}=6,0$~\r{A}, $U_{\rm s}= +2.60$~V, $I_{\rm t}=80$~pA.}
\end{figure}

\begin{figure}
\includegraphics[width=0.215\textwidth]{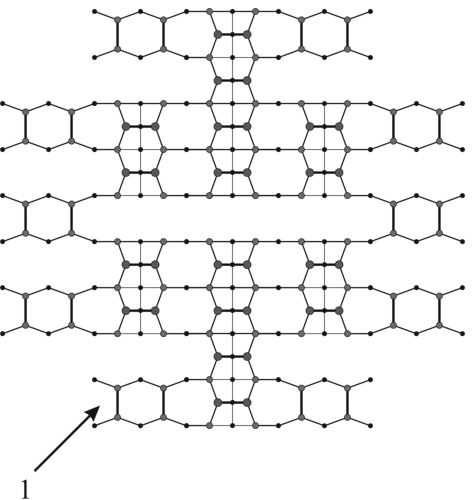}(a)
\includegraphics[width=0.215\textwidth]{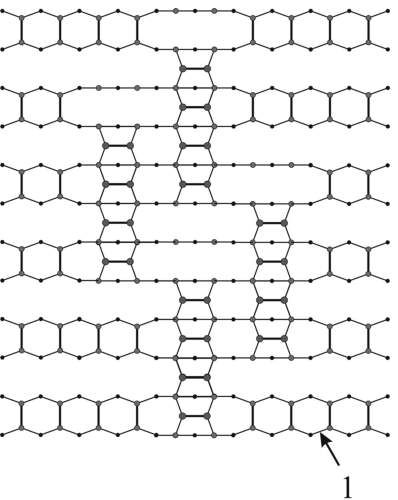}(b)
\includegraphics[width=0.215\textwidth]{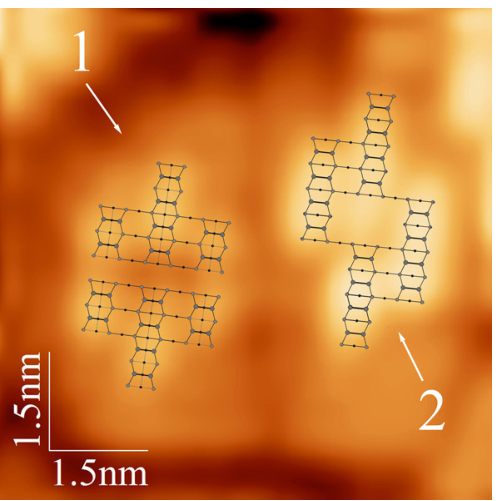}(c)
\caption{\label{fig:models}Models of nuclei of Ge hut clusters corresponding to the images given in Fig.~\ref{fig:nuclei}(b): (a) a pyramid,     
(b) a  wedge  [1 is the wetting layer in the plots (a) and (b)]; (c) the models superimposed on the image given in Fig.~\ref{fig:nuclei}(b), the numbering is the same as in Fig.~\ref{fig:nuclei}(b).
}
\end{figure}

\begin{figure}
\includegraphics[width=0.4\textwidth]{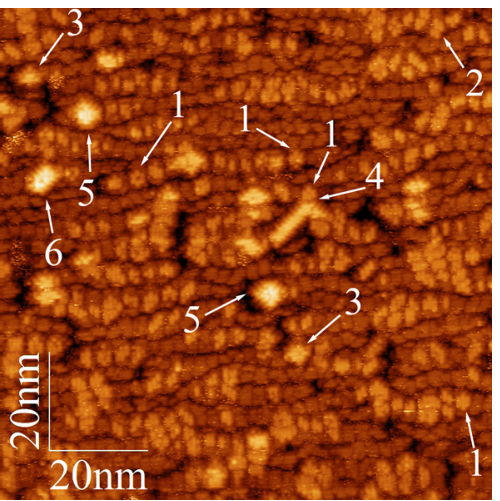}(a)
\includegraphics[width=0.4\textwidth]{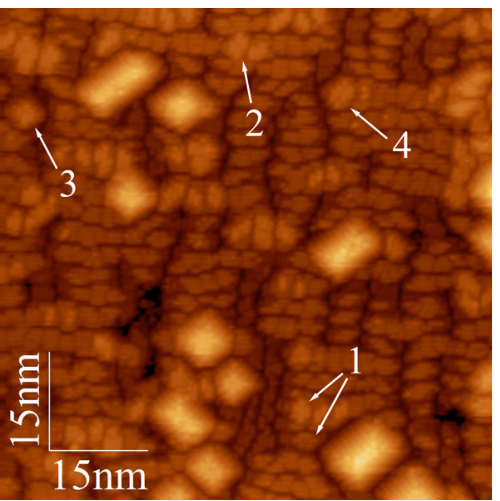}(b)
\caption{\label{fig:dots}
STM image of  Ge wetting layer on Si(001): (a)  $h_{\rm Ge} = 5.4$~\r{A}  ($U_{\rm s}=+1.80$~V, $I_{\rm t}=100$~pA)  and (b) $h_{\rm Ge} = 6.0$~\r{A}   
($U_{\rm s}=+2.50$~V, $I_{\rm t}=80$~pA). Examples of characteristic features are numbered as follows:  nuclei of pyramids (1) and  wedges  (2) [1 ML high over WL],  small pyramids (3) and wedges (2) [2 ML high over WL, a $\rm \Gamma$-like wedge \cite{classification} is observed in the image (a)],  3 ML high pyramids (5) and   wedges (6).}
\end{figure}

Investigating an evolution of the hut arrays we have arrived at a conclusion that a moment of an array nucleation during MBE precedes a moment of formation of the first hut on the WL.\footnote{Or, in other words, it foreruns a moment of formation of the first \{105\} faceted cluster with the height-to-width ratio of 1:10 on WL.} It is not a paradox. Hut cluster arrays nucleate when the first hut nuclei arise on the $(M\times N)$ patch of the wetting layer. This process is illustrated by Fig.~\ref{fig:nucleation}. An image (a) demonstrates a typical STM micrograph of the WL with the $(M\times N)$ patched structure ($h_{\rm Ge}=4.4$~\r{A}). This image does not demonstrate any feature which might be interpreted as a hut nucleus \cite{Hut_nucleation}. Such features first arise at the  coverages $\sim 5$\,\r{A}: they are clearly seen in the image (b), which demonstrates a moment of the array birth ($h_{\rm Ge}=5.1$~\r{A}), and numbered by `1' for the pyramid nucleus and `2' for the wedge one (several analogous formations can be easily found by the readers on different patches). However, no hut clusters are seen in this picture. 

Our interpretation is based on the results reported in Ref.~\cite{Hut_nucleation} which evidenced that there are two different types of nuclei on Ge wetting layer which evolve in the process of Ge deposition to pyramidal and wedge-like hut clusters. 
  Having assumed that nuclei emerge on WL as combinations of dimer pairs and/or longer chains of dimers in epitaxial configuration \cite{epinucleation} and correspond to the known structure of apexes specific for each hut species \cite{classification,atomic_structure}
we have investigated WL patches, 1 monolayer (ML) high formations on them and clusters of different heights (number of steps) over WL.  As a result, we succeeded to select two types of formations different in symmetry and satisfying the above requirements, which first appear  at a coverage of $\sim 5$~\r{A} and then arise on  WL during the array growth. We have interpreted them as hut nuclei, despite their sizes  are much less than those predicted by the first principle calculations \cite{hut_stability}, and traced their evolution to huts.

The nuclei formation is illustrated by Fig.~\ref{fig:nuclei}. The surface structure of the  $(M\times N)$ patches is shown in the micrograph (a). The letter `$c$' indicates the 
$c(4\times 2)$ reconstructed patch, `$p$' shows  a patch with the $p(2\times 2)$  reconstruction \cite{Iwawaki_initial,Iwawaki_dimers}. Both reconstructions are always detected simultaneously that means they are very close (or degenerate) by energy. The image (b) shows two adjacent patches reconstructed by the born nuclei: `1' and `2' denote the pyramid (a formation  resembling a blossom) and wedge nuclei respectively \cite{Hut_nucleation}. Their structural models derived from  many STM images \cite{classification,atomic_structure,Hut_nucleation} are presented in Fig.~\ref{fig:models}(a, b) and superimposed on the images of the nuclei in  Fig.~\ref{fig:models}(c). Note that both types of nuclei arise at the same moment of the MBE growth. It means that they are degenerate by the formation energy. An issue why two different structures, rather than one, arise  to  relief the WL strain   remains open, however. 

It is necessary to remark here that the nuclei are always observed to arise on sufficiently large WL patches. There must be enough room for a nucleus on a single patch. A nucleus cannot be housed on more than one patch. So, cluster nucleation is impossible on little (too narrow or short) patches (Fig.~\ref{fig:nuclei}(b)).

The hut nucleation goes on during the array further evolution. Fig.~\ref{fig:dots}
illustrates this process. An array shown in Fig.~\ref{fig:dots}(a) ($h_{\rm Ge} = 5.4$~\r{A}) consists of 1-ML nuclei (`1' and `2'), 2-ML and 3-ML pyramids and wedges (`3' and `5', `4' and `6' mark pyramids and wedges respectively).\footnote{Hereinafter, the cluster heighs are counted from the WL top.}  
Fig.~\ref{fig:dots}(b) ($h_{\rm Ge} = 6.0$~\r{A}) demonstrates the simultaneous presence of nuclei (`1' and `2') and 2-ML huts (`3' and `4') with the growing much higher clusters.  

Hut cluster nucleation on the WL surface continues until the final phase of the array life. This peculiarity distinguishes low-temperature growth mode from the high-temperature one \cite{classification}.

\subsection{\label{sec:structure}Structural models}

\begin{figure}
\includegraphics[width=0.33\textwidth]{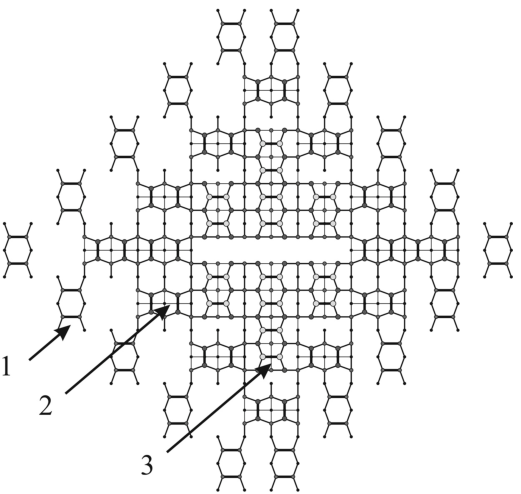}(a)
\includegraphics[width=0.45\textwidth]{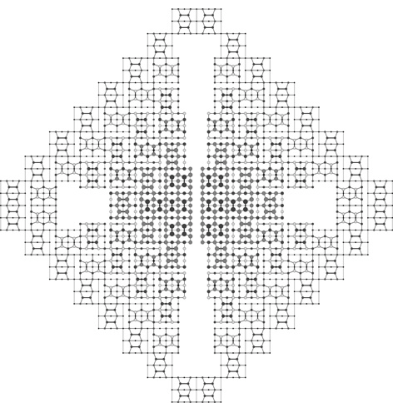}(b)
\caption{\label{fig:pyramids} Top views of the pyramidal QDs consisting of (a) 2  and (b) 6 monoatomic steps and  (001) terraces on the wetting layer (1, 2 and 3 designate wetting layer, the first and the second layers of the clusters respectively). }
\end{figure}

\begin{figure}
\includegraphics[width=0.21\textwidth]{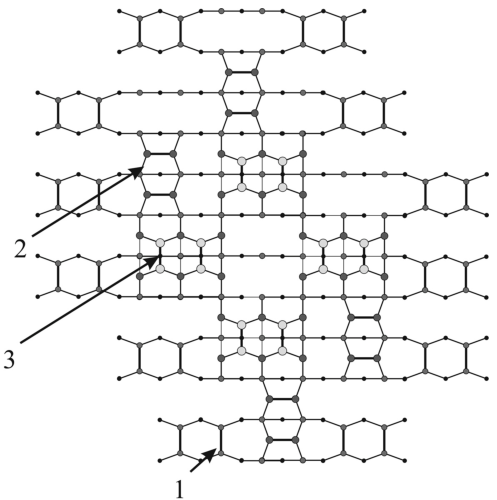}(a) 
\includegraphics[width=0.30\textwidth]{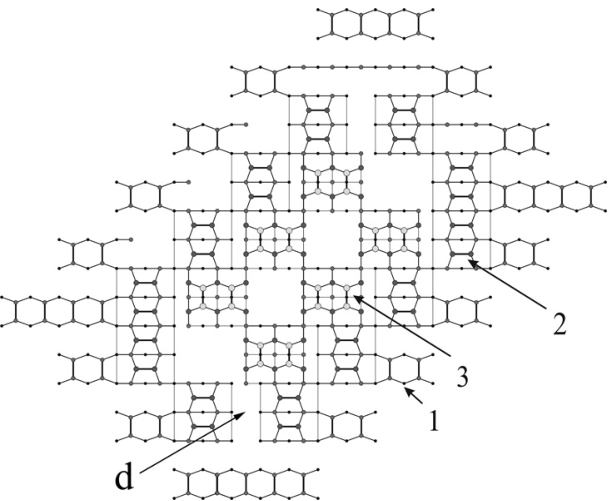}(b)
\includegraphics[width=0.4\textwidth]{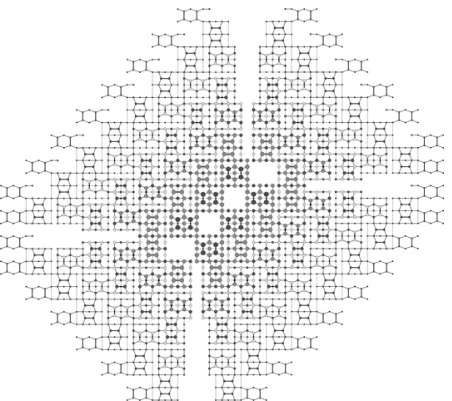}(c)
\caption{\label{fig:wedges} Growth of a wedge-like cluster: 
(a) reconstruction of the first layer  of  a forming wedge  during addition of epi-oriented dimer pairs of the second (001) terrace; 
plots of atomic structures of a Ge  wedge-shaped hut clusters composed by (b) 2 and (c)  6 monoatomic steps and  (001) terraces on the wetting layer  (the numbering is the same as in Fig.~\ref{fig:pyramids};  d marks a defect arisen because of one translation uncertainty of the left dimer pair position).
}
\end{figure}

It is commonly adopted that the hut clusters grow by successive filling the (001) terraces of the $\{105\}$ faces by the dimer rows \cite{Kastner}. However, formation of the  sets of steps and terraces requires the hut base sides to be parallel to the  $<$100$>$ directions. The pyramid nucleus satisfies this requirement, its sides aline with \textless 100\textgreater. Thus the pyramids  grow  without phase transition when the second and subsequent layers are added (Fig.~\ref{fig:pyramids}). Only nucleus-like structures of their apexes are rotated $90^{\circ}$ with respect to the rows on previous terraces to form the correct epitaxial configuration when the heights are increased by 1\,ML, but this rotation does not violate the symmetry of  the previous layers of the cluster.

A different scenario of growth  of the wedge-like clusters have been observed. Two base sides the wedge nucleus does not aline with $<$100$>$  (Fig.~\ref{fig:models}(b)). The ridge structure of a wedge is  different from the nucleus structure presented in Fig.~\ref{fig:models}(b) \cite{classification,Hut_nucleation,atomic_structure}. It was shown in Ref.~\cite{Hut_nucleation} that the structure of the wedge-like cluster arise due to rearrangement of  rows of the first layer in the process of the second layer formation (Fig.~\ref{fig:wedges}(a)).  
The phase transition in the first layer  generates the base  with all sides directed along the $<$100$>$ axes which is necessary to give rise to the $\{105\}$ faceted cluster. 
After the transition, the elongation of the elementary structure is possible only along a single axis which is determined by the symmetry  (along the arrows in Fig.~\ref{fig:wedges}(a)). A formed 2-ML wedge is plotted in Fig.~\ref{fig:wedges}(b). A structure of the 6-ML wedge appeared as a result of further in-height growth is  shown in  
Fig.~\ref{fig:wedges}(c). The ridge structures of the 2-ML and 6-ML wedges is seen to coincide, which is not the case for different cluster heights. A complete set of the wedge ridges for different cluster heights can be obtained by filling the terraces by epi-oriented pairs of dimers.

It should be noted also that according to the proposed model the wedge-like clusters always contain point defects on the triangular (short) facets. The defects are located in the upper corners of the facets  and caused by uncertainty of one translation in the position a dimer pair  which forms the penultimate terrace of the triangular facet (Fig.~\ref{fig:wedges}). The predicted presence of these defects removes the degeneracy of the facets and hence 
an issue of the pyramid symmetry violation which occurs if the pyramid-to-wedge transition is assumed (this issue was discussed in detail in Ref.~\cite{classification}). In addition,
the vacancy-type defects may decrease the energy of addition of new atoms  to the triangular facets and stimulate the quicker growth on them than on the trapezoidal ones and rapid elongation of wedges.
These defects are absent on the facets of the pyramidal huts. 
Their triangular facets  are degenerate. Therefore, as it follows from our model, the trapezoidal and triangular facets of the wedge 
are not degenerate with respect to one another even at very beginning of cluster growth. The wedges can easily elongate by growing on the 
triangular facets faster than on trapezoidal ones. Pyramids, having degenerate facets, cannot elongate and grow 
only in height outrunning wedges. This explains greater heights of pyramids \cite{classification}.

Analyzing the deduced structural models of pyramids and wedges, as well as their behaviour during the array nucleation and growth, we have come to conclusion that shape transitions between the clusters of different species are prohibited \cite{classification,Hut_nucleation,atomic_structure}.

\subsection{\label{sec:facets}Facets}

\begin{figure}[h]
\includegraphics[width=0.49\textwidth]{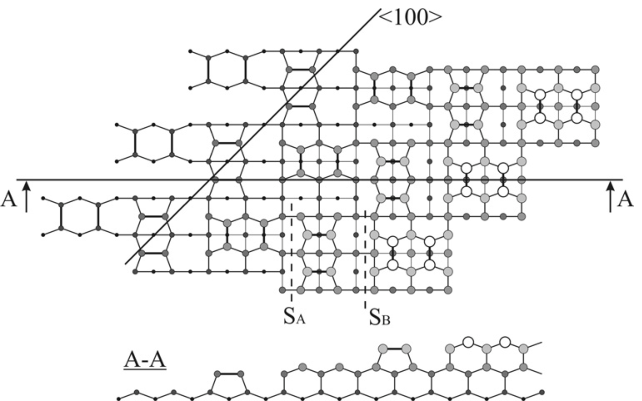}(a)
\includegraphics[width=0.3\textwidth]{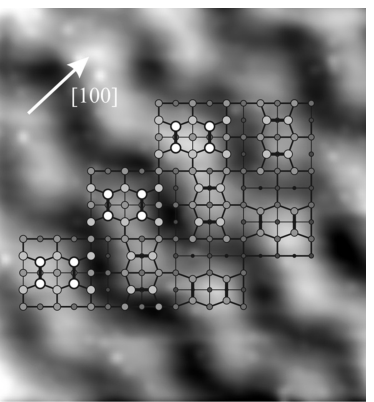}(b)
\caption{\label{fig:face}
(a) A structural model of the $\{105\}$ facet of hut clusters  derived from the plots given in Figs.~\ref{fig:pyramids} and~\ref{fig:wedges} corresponds to the PD (pairs of dimers) model \cite{Mo},  S$_{\rm A}$ and  S$_{\rm B}$ are commonly adopted designations of the monoatomic steps \cite{Chadi}, atoms situated on higher terraces are shown by larger circles.
(b) The schematic of the  facet  superimposed on its STM image  ($4.3\times 4.4$~nm, $U_{\rm s}= +3.0$~V, $I_{\rm t}=100$~pA), the [100] direction is parallel to the corresponding base side, the steps rise from the lower right to the upper left corner.
}
\end{figure}

The presented models  allowed us to 
deduce a structure of the $\{105\}$ facets (Fig.~\ref{fig:face}(a)). This model resulting from the above simple crystallographic consideration corresponds to the paired dimers  (PD) \cite{Mo}  rather than more recent rebonded step  (RS) model \cite{Fujikawa,Facet-105} which is now believed to improve the previous PD model by Mo {\it et al}.

\begin{figure}
\includegraphics[width=0.215\textwidth]{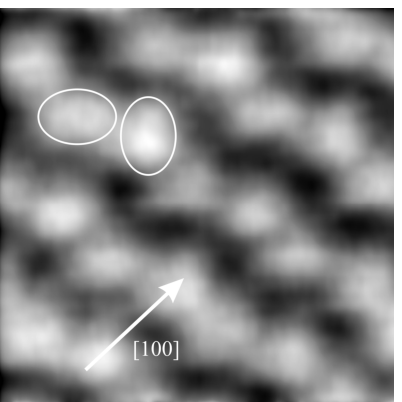}(a)
\includegraphics[width=0.215\textwidth]{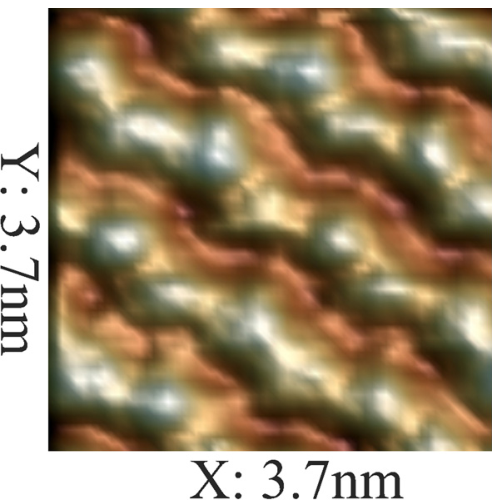}(b)
\caption{\label{fig:facet}
(a) 2D  and   (b) 3D  STM images  of the same area on Ge hut cluster facet ($h_{\rm Ge}=10$~\r{A}, $T_{\rm gr}=360^{\circ}$C,
$U_{\rm s}=+2.1$~V, $I_{\rm t}=80$~pA). The sides of the cluster base lie along the [100] direction; structural units revealed on the free surfaces of the (001) terraces and interpreted as paired dimers are marked out. }
\end{figure}

A direct STM exploration of the $\{105\}$ facets confirms the derived model.
Being superposed with the empty state STM image of the cluster  $\{105\}$ facet it demonstrates an excellent agreement with the experiment (Fig.~\ref{fig:face}(b)).
A typical STM image of the QD facet is presented in Fig.~\ref{fig:facet}. Characteristic distances on the facets are as follows: $\sim 10.5$~\r{A} in the \textless 100\textgreater~directions (along the corresponding side of the base) and  $\sim 14$~\r{A} in the normal (\textless 051\textgreater) directions.
The facets are composed by structural units which are outlined by ellipses in Fig.~\ref{fig:facet}(a) and can be arranged along either [110] or [1${\overline 1}$0] direction on the (001) plane. We have interpreted them as pairs of dimers. Their positional relationship is  obviously seen in the 3D micrograph presented in Fig.~\ref{fig:facet}(b).
 
Dangling bonds of the derived  $\{105\}$-PD facets, due to high chemical activity, 
 may stimulate Ge atom addition   and cluster growth. Thus
less stability and higher activity  of the $\{105\}$-PD  facets compared to the \linebreak 
Ge(105)/Si(105)-RS plane, which is usually adopted in the literature  for simulation of hut  $\{105\}$ facets, may cause fast completion of hut terraces during epitaxy and be responsible (or even be necessary) for hut formation and growth.

\subsection{\label{sec:density}Cluster density and fractions}

\begin{figure}
\includegraphics[width=0.215\textwidth]{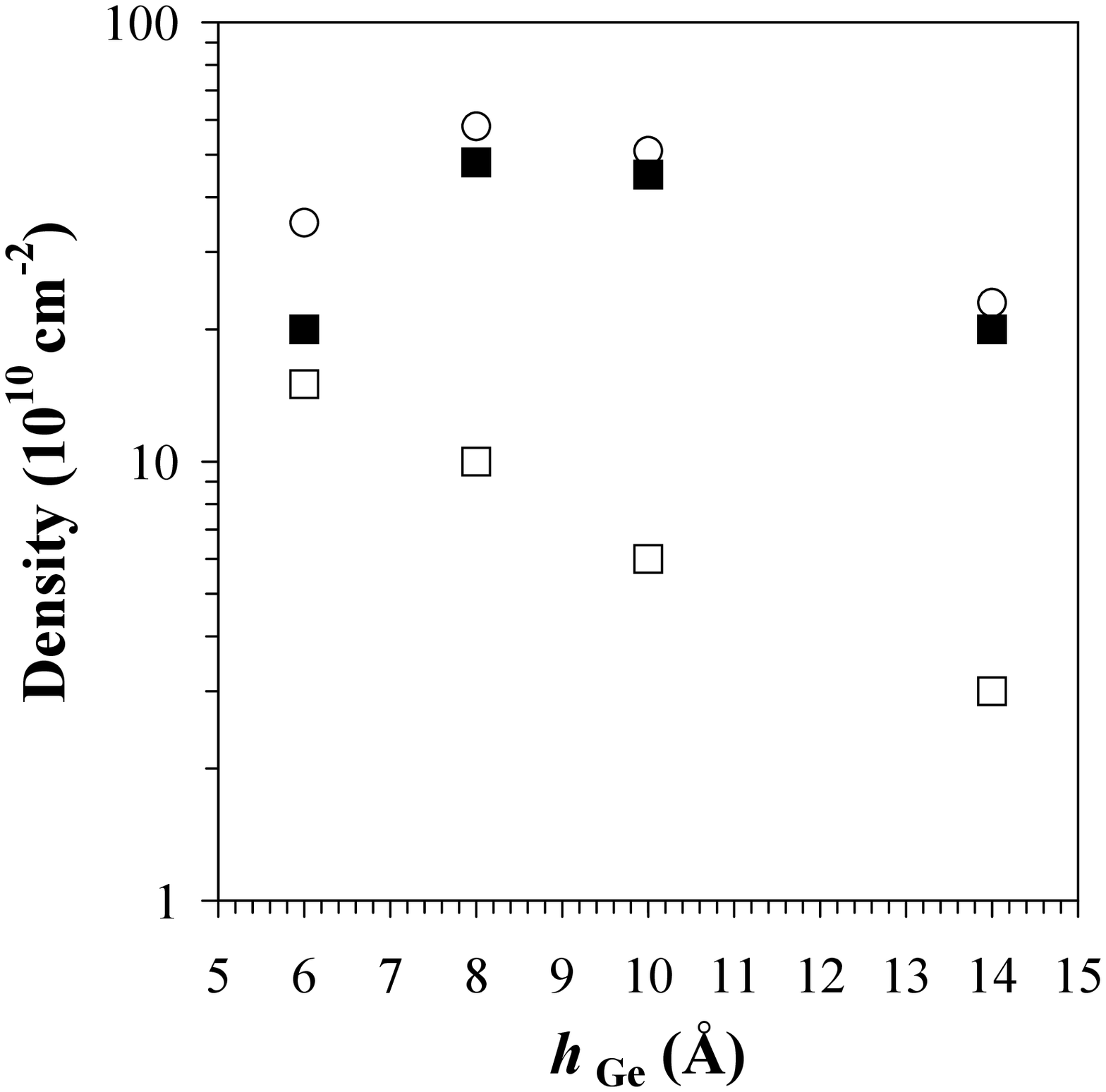}(a)
\includegraphics[width=0.215\textwidth]{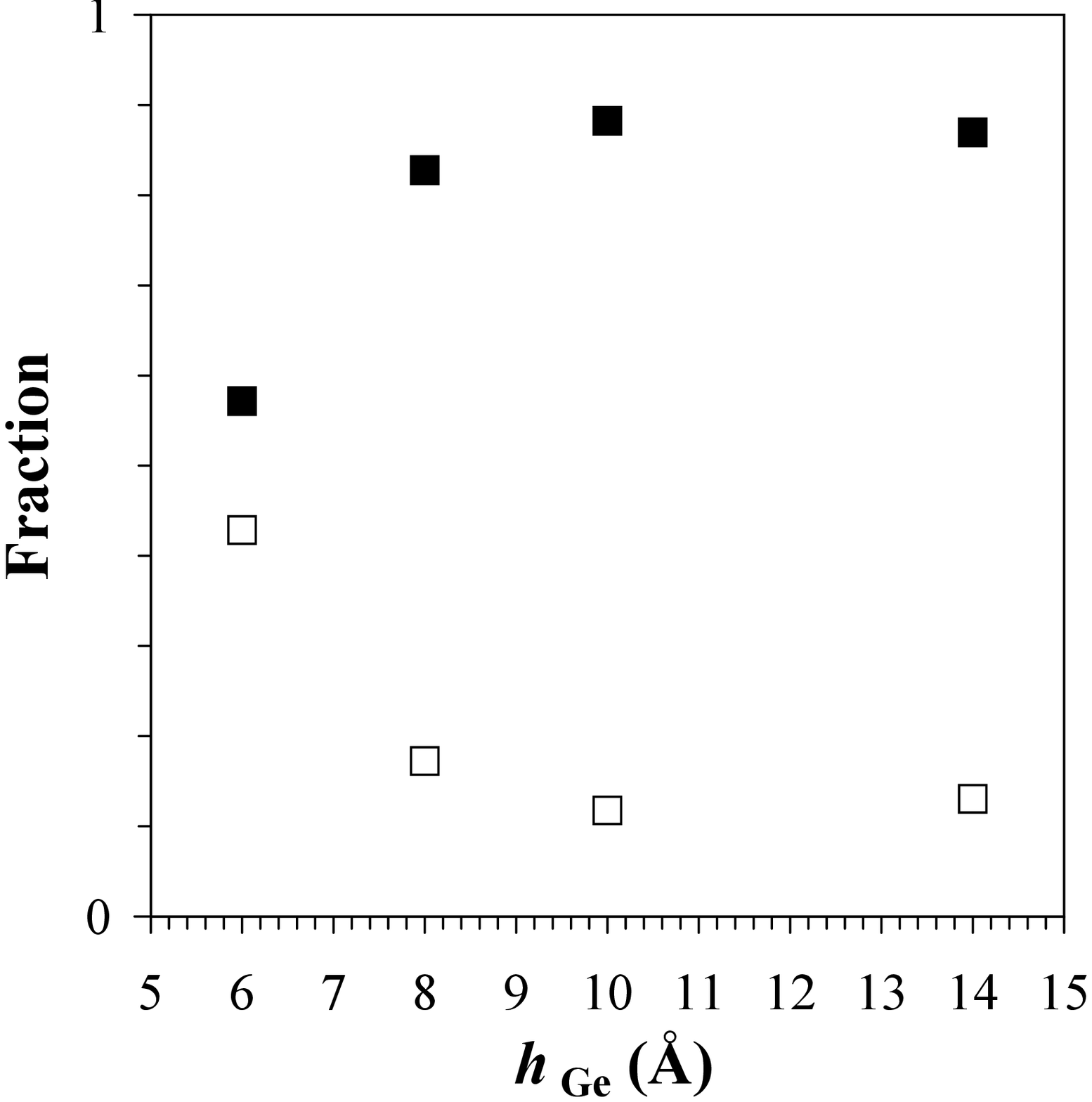}(b)
\caption{\label{fig:density} (a)
 Density  and (b) fraction  of the Ge clusters in the arrays  formed at  $T_{\rm gr} = 360^\circ$C ($\square$~marks the pyramids, $\blacksquare$~designates the wedges,  $\Circle$~is the total density).}
\end{figure}

Fig.~\ref{fig:density}(a) plots the dependence of the cluster density on $h_{\rm Ge}$ for different clusters in the arrays. It is seen that  the density of wedges rises starting from $D_{\rm w}\approx 1,8\times 10^{11}$~cm$^{-2}$ at the beginning of the three-dimensional growth of Ge (the estimate is obtained by data extrapolation to $h_{\rm Ge}= 5$~\r{A}) and  reaches the maximum of $\sim 5\times 10^{11}$~cm$^{-2}$ at $h_{\rm Ge} \sim 8$~\r{A}, the total  density of clusters at this point $D_{\rm \Sigma} \sim 6\times 10^{11}$~cm$^{-2}$ is also maximum. Then both  $D_{\rm w}$ and $D_{\rm \Sigma}$ slowly go down until the two-dimensional growth of Ge starts at $h_{\rm Ge} \sim 14$~\r{A} and  $D_{\rm \Sigma}\approx D_{\rm w} \sim 2\times 10^{11}$~cm$^{-2}$ (the contribution of pyramids $D_{\rm p}$ to  $D_{\rm \Sigma}$ becomes negligible---~$\sim 3\times 10^{10}$~cm$^{-2}$---at this value of $h_{\rm Ge}$). 
 The pyramid density exponentially drops  as the value of $h_{\rm Ge}$ grows ($D_{\rm p} \approx 5\times 10^{11} \exp\{-2.0\times 10^7\,h_{\rm Ge}\}$, $h_{\rm Ge}$ is measured in centimeters). The maximum value of $D_{\rm p} \approx 1.8\times 10^{11}$~cm$^{-2}$ obtained from extrapolation to $h_{\rm Ge} = 5$~\r{A} coincides with the estimated initial value of $D_{\rm w}$.

The graphs of cluster fractions in the arrays versus $h_{\rm Ge}$ are presented in Fig.~\ref{fig:density}(b). Portions of pyramids and wedges initially very close ($\sim 50\%$ at $h_{\rm Ge}\sim 5$~\AA) rapidly become different as $h_{\rm Ge}$ rises. The content of pyramids monotonically falls. The fraction of the wedge-like clusters is approximately $57\%$ at the early stage of the array growth ($h_{\rm Ge} = 6$~\r{A}) and becomes $82\%$ at $h_{\rm Ge} = 8$~\r{A}. At further growth of the array, the content of the wedges reaches the saturation at the level of approximately $88\%$ at $h_{\rm Ge} = 10$~\r{A}. 

The inference may be made from this observation that contrary to the intuitively expected from the consideration of symmetry, the wedge-like shape of the clusters is energetically more advantageous than the pyramidal one, and the more advantageous the more Ge atoms (and the more the number of terraces) constitute the cluster. 
The probability of nucleation appears to be close to 1/2 for both wedges-like and pyramidal clusters at the initial stage of the array formation and low growth temperatures. Then, as the array grows, the formation of pyramids becomes hardly probable and most of them, which have  already  formed, vanish whereas the nucleation and further growth of wedges continues. The Ge pyramids on the Si(001) surface turned out to be less stable objects than the wedges. 

Notice also that at $T_{\rm gr} = 360^\circ$C and the flux of Ge atoms ${\rm d} h_{\rm Ge}/{\rm d}t = 0.15$~\r{A}/s, the point $h_{\rm Ge} = 10$~\r{A} is particular. Not only the fraction of pyramids saturates at this point but the array in whole has the most uniform sizes of the clusters composing it (Figs.~\ref{fig:dots}, \ref{fig:arrays} and \ref{fig:finish}). This is concluded by us not only on the basis of analysis of the STM images of the Ge/Si(001) arrays but also from the data of the Raman  scattering  by the Ge/Si heterostructures with different low-temperature arrays of Ge quantum dots \cite{our_Raman_en,Raman_conf}. We refer to such arrays as optimal.

\subsection{\label{sec:life}Array life cycle}

\begin{figure}
\includegraphics[width=0.215\textwidth]{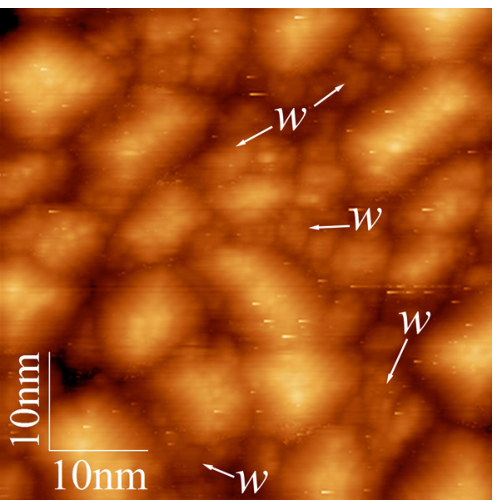}(a)
\includegraphics[width=0.215\textwidth]{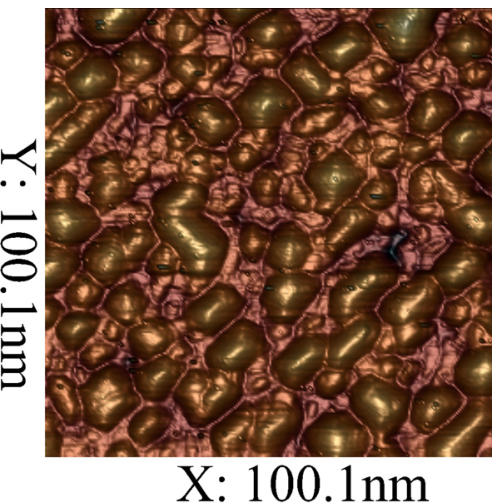}(b)
\includegraphics[width=0.215\textwidth]{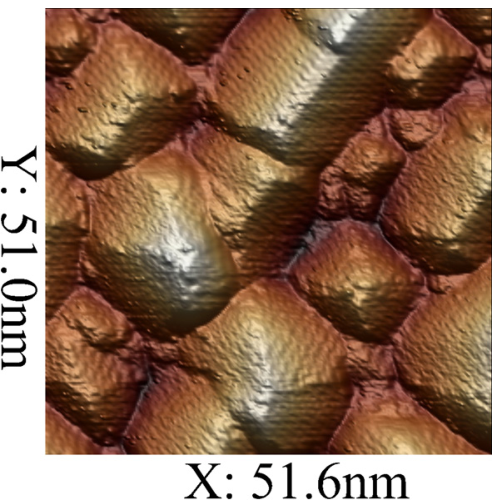}(c)
\includegraphics[width=0.215\textwidth]{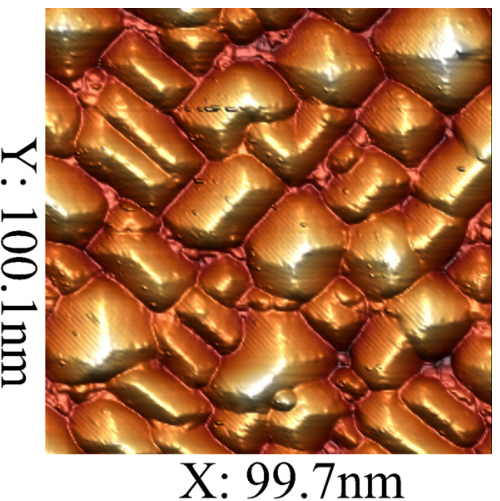}(d)
\caption{\label{fig:arrays} 
 STM 2D and 3D micrographs of Ge hut cluster dense arrays at different coverages ($T_{\rm gr} = 360^\circ$C): (a),\,(b)  $h_{\rm Ge}=8$~\r{A} [(a)~$50.6\times 49.9$\,nm, $w$ is the wetting layer, $U_{\rm s}=+2.0$~V, $I_{\rm t}=80$~pA; (b) $U_{\rm s}=+2.0$~V, $I_{\rm t}=100$~pA]; (c),\,(d) $h_{\rm Ge}=10$~\r{A} [(c),\,(d) $U_{\rm s}=+2.1$~V, $I_{\rm t}=100$~pA].}
\end{figure}

\begin{figure}
\includegraphics[width=0.215\textwidth]{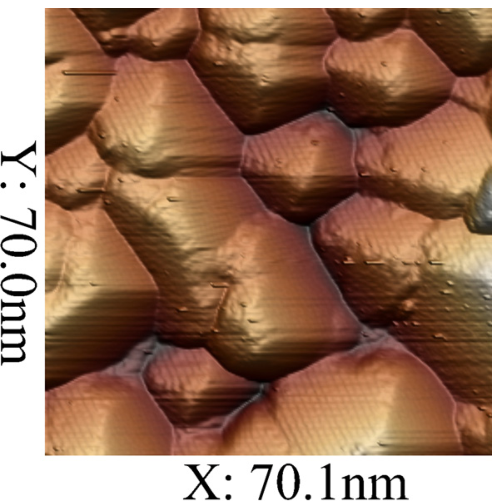}(a)
\includegraphics[width=0.215\textwidth]{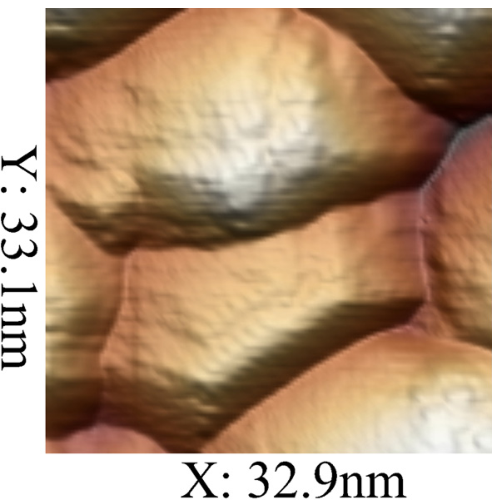}(b)
\includegraphics[width=0.215\textwidth]{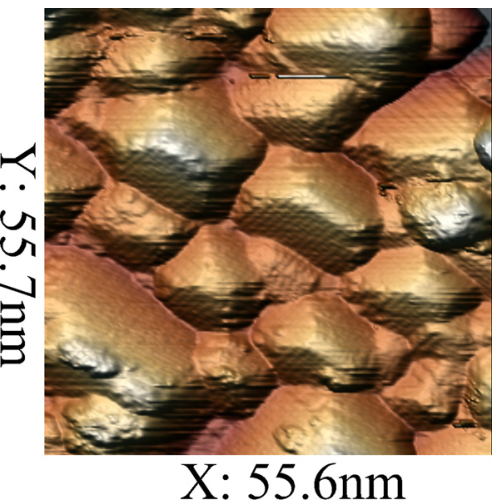}(c)
\includegraphics[width=0.215\textwidth]{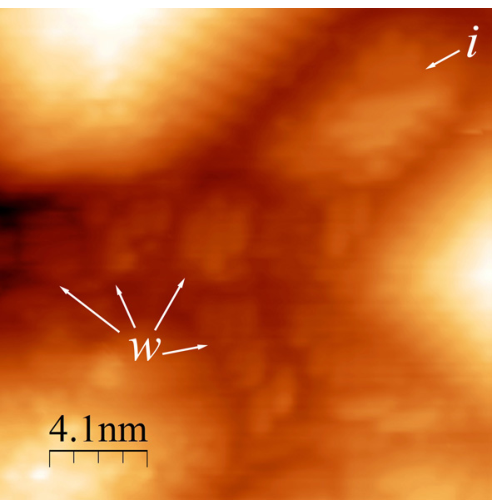}(d)
\caption{\label{fig:finish} 
 STM topographs of Ge hut cluster dense arrays at different coverages ($T_{\rm gr} = 360^\circ$C): (a),\,(b)  $h_{\rm Ge}=14$\,\r{A} [(a) $U_{\rm s}=+1.75$\,V, $I_{\rm t}=80$\,pA, (b) $U_{\rm s}=+3.0$\,V, $I_{\rm t}=100$\,pA]; (c),\,(d) $h_{\rm Ge}=15$\,\r{A} [(c) $U_{\rm s}=+2.0$\,V, $I_{\rm t}=120$\,pA,  (d)~$20.3\times 20.4$\,nm, $U_{\rm s}=+3.6$\,V, $I_{\rm t}=120$\,pA], $w$ indicates the wetting layer patches, $i$ shows a distorted small Ge island  3\,ML high over WL.}
\end{figure}

A qualitative model accounting for the presence of the particular point at the low-temperature array growth is simple. The case is that at low enough temperatures of the array growth, the new Ge cluster nucleation competes with the process of growth of earlier formed clusters. The height of the dominating wedge-like clusters is observed to be limited by some value depending on $T_{\rm gr}$.\footnote{Note that we did not observe a height limitation of pyramids. We suppose that they may give rise to one of the types of  array defects---huge clusters---which sometimes appear among ordinary huts \cite{defects_ICDS-25}.} At small $h_{\rm Ge}$, Ge clusters are small enough and the distances between them are large enough compared to the Ge atom (or dimer) diffusion (migration) length on the surface for nucleation of new clusters on the Ge wetting layer in the space between the clusters (Figs.~\ref{fig:dots}, ~\ref{fig:arrays}(a),(b)). At $h_{\rm Ge} = 10$~\r{A} and the above ${\rm d} h_{\rm Ge}/{\rm d}t$  values, the equilibrium of parameters (cluster sizes and distances between them, diffusion length at given temperature, Ge deposition rate, etc.) sets in, the rate of new cluster nucleation is decreased and the abundant Ge atoms are mainly spent to the growth of the available clusters (Fig.~\ref{fig:arrays}(c),(d)). After the clusters reach their height limit and in spite of it, Ge atoms continue to form up their facets. As soon as most of the clusters reach the height limit, nucleation of new clusters becomes energetically advantageous again and the nucleation rate rises. The second phase of clusters appears on the wetting layer and fills whole its free surface as $h_{\rm Ge}$ is increased (Fig.~\ref{fig:finish}). Further increase of $h_{\rm Ge}$ results in two-dimensional growth mode. It is clear now why the array is the most homogeneous (optimal) at $T_{\rm gr} = 360^\circ$C and $h_{\rm Ge} = 10$~\r{A} whereas the dispersion of the cluster sizes is increased at higher and lower values of $h_{\rm Ge}$ because of the small clusters containing in the array. It is clear also that the optimal array will appear at different value of $h_{\rm Ge}$ when $T_{\rm gr}$ or ${\rm d} h_{\rm Ge}/{\rm d}t$ are different.

As it follows from the data presented in this section and Section~\ref{sec:formation}  the Ge hut array evolution and life cycle goes through three main phases: at  $T_{\rm gr}=360^{\circ}$C, the array nucleates at   $h_{\rm Ge} \sim 5$\,\r{A} (Fig.~\ref{fig:nucleation}), it reaches ripeness and optimum to  $h_{\rm Ge} \sim 10$\,\r{A} (Fig.~\ref{fig:arrays}) and finishes its evolution at $h_{\rm Ge} \sim 14$\,\r{A} by filling whole the surface (Fig.~\ref{fig:finish}). Most of clusters start coalescing (Fig.~\ref{fig:finish}(b)) and 2D growth begins at greater  $h_{\rm Ge}$ (Fig.~\ref{fig:finish}(c)). 

Nevertheless, free areas of WL still remain even at  $h_{\rm Ge} = 15$\,\r{A} (Fig.~\ref{fig:finish}(d)). The structure of the parches (`$w$') stays the same as in the beginning of the array formation although the WL regions are  surrounded by large huts. Small 3D islands (`$i$'), although very distorted, are still recognizable on WL between the large huts. The hut nucleation on WL goes on even at as high coverages as 15\,\r{A} when virtually total coalescence of the mature huts  have already happened.

\section{Conclusion}
\label{sec:conclusion}

In summary,  we have studied the array nucleation phase and identified  the nuclei of both hut species, determined their atomic structure and observed the moment of appearance of the first generation of the nuclei on WL. We have investigated with high spatial resolution the peculiarities of each species of huts and their growth and derived their atomic structures. We have concluded that the wedge-like huts form due to a phase transition reconstructing the first atomic step of the growing cluster when dimer pairs of its second atomic layer stack up; the pyramids grow without  phase transitions. In addition, we have come to conclusion  that wedges contain  vacancy-type defects on the penultimate terraces of their triangular facets which may decrease the energy of addition of new atoms  to these facets and stimulate the quicker growth on them than on the trapezoidal ones and rapid elongation of wedges. We have shown also comparing the structures and growth of pyramids and wedges that shape transitions between them are impossible. And finally, we have explored the array evolution during MBE right up to the concluding phase of its life  when most clusters coalesce and start forming a  nanocrystalline 2D layer.



%
%

\bibliography{QD-arrays-EMRS}

\end{document}